\documentclass[sn-mathphys,Numbered]{sn-jnl}
\usepackage[markup=underlined]{changes}
\usepackage{todonotes}
\setcommentmarkup{\todo[color={authorcolor!20},size=\scriptsize]{#3: #1}}

\usepackage{graphicx}%
\usepackage{multirow}%
\usepackage{amsmath,amssymb,amsfonts}%
\usepackage{amsthm}%
\usepackage{mathrsfs}%
\usepackage[title]{appendix}%
\usepackage{xcolor}%
\usepackage{textcomp}%
\usepackage{manyfoot}%
\usepackage{booktabs}%
\usepackage{algorithm}%
\usepackage{algorithmicx}%
\usepackage{algpseudocode}%
\usepackage{hyperref}       %
\usepackage[utf8]{inputenc} %
\usepackage[T1]{fontenc}    %
\usepackage{url}            %
\usepackage{booktabs}       %
\usepackage{nicefrac}       %
\usepackage{microtype}      %
\usepackage{xcolor}         %
\usepackage[export]{adjustbox}
\usepackage{wrapfig,amsthm,textcomp,amssymb}
\usepackage{colortbl}
\usepackage{pifont}%
\usepackage[capitalise,noabbrev,nameinlink]{cleveref}
\usepackage{adjustbox}
\usepackage{subfigure}
\usepackage{soul}
\usepackage{cancel}
\usepackage{graphicx}
\usepackage{listings}%
\definecolor{good}{HTML}{ffd2ad}
\definecolor{good}{HTML}{74c476}
\begin{document}

\title[Article Title]{Geometry-enhanced Pre-training on Interatomic Potentials}

\author[1,2]{\fnm{Taoyong} \sur{Cui}}
\equalcont{These authors contributed equally to this work. This work was done during their internship at Shanghai Artificial Intelligence Laboratory.}
\author[1,3,4]{\fnm{Chenyu} \sur{Tang}}
\equalcont{These authors contributed equally to this work. This work was done during their internship at Shanghai Artificial Intelligence Laboratory.}
\author*[1]{\fnm{Mao} \sur{Su}}\email{sumao@pjlab.org.cn}
\author*[1]{\fnm{Shufei} \sur{Zhang}}\email{zhangshufei@pjlab.org.cn}
\author[1]{\fnm{Yuqiang} \sur{Li}}
\author[1]{\fnm{Lei} \sur{Bai}}
\author[2]{\fnm{Yuhan} \sur{Dong}}
\author[5,6]{\fnm{Xingao} \sur{Gong}}
\author[1]{\fnm{Wanli} \sur{Ouyang}}

\affil[1]{\orgname{Shanghai Artificial Intelligence Laboratory}, \orgaddress{\city{Shanghai}, \postcode{200232}, \country{China}}}

\affil[2]{\orgdiv{Shenzhen International Graduate School}, \orgname{Tsinghua University}, \orgaddress{\city{Shenzhen}, \postcode{518055}, \country{China}}}

\affil[3]{\orgdiv{CAS Key Laboratory of Theoretical Physics}, \orgname{Institute of Theoretical Physics}, \orgname{Chinese Academy of Sciences}, \orgaddress{\city{Beijing}, \postcode{100190}, \country{China}}}

\affil[4]{\orgdiv{School of Physical Sciences}, \orgname{University of Chinese Academy of Sciences}, \orgaddress{\city{Beijing}, \postcode{100049}, \country{China}}}

\affil[5]{\orgdiv{Key Laboratory for Computational Physical Sciences (MOE), State Key Laboratory of Surface Physics, Department of Physics}, \orgname{Fudan University}, \orgaddress{\city{Shanghai}, \postcode{200433}, \country{China}}}

\affil[6]{\orgname{Shanghai  Qi Zhi Institute}, \orgaddress{\city{Shanghai}, \postcode{200232}, \country{China}}}


\abstract{Machine learning interatomic potentials (MLIPs) describe the interactions between atoms in materials and molecules by learning from a reference database generated by ab initio calculations. MLIPs can accurately and efficiently predict the interactions and have been applied to various fields of physical science. However, high-performance MLIPs rely on a large amount of labeled data, which are costly to obtain by ab initio calculations. Compared with ab initio methods, empirical interatomic potentials enable classical molecular dynamics (CMD) simulations that generate atomic configurations efficiently, but their energies and forces are not accurate, which can be regarded as non-labeled for MLIPs. In this paper, we propose a geometric structure learning framework that leverages the unlabeled configurations to improve the performance of MLIPs. Our framework consists of two stages: firstly, using CMD simulations to generate unlabeled configurations of the target molecular system; and secondly, applying geometry-enhanced self-supervised learning techniques, including masking, denoising, and contrastive learning, to capture structural information. We evaluate our framework on various benchmarks ranging from small molecule datasets to complex periodic molecular systems with more types of elements. We show that our method significantly improves the accuracy and generalization of MLIPs with only a few additional computational costs and is compatible with different invariant or equivariant graph neural network architectures. Our method enhances MLIPs and advances the simulations of molecular systems.}


\maketitle

\section{Introduction}\label{sec1}

Molecular Dynamics (MD) simulation provides atomic insights in many areas such as physics, chemistry, biology, and material science \cite{hospital2015molecular, senftle2016reaxff, karplus1990molecular, yao_applying_2022}. The accuracy and efficiency of MD simulations depend on the choice of interatomic potentials, which are mathematical functions that describe the potential energy of atoms in a molecular system. In classical MD (CMD) simulations, the interatomic potentials are modeled by empirical formula with parameters to be fitted \cite{kaminski2001evaluation}. The CMD simulations are computationally efficient but inaccurate for many applications, especially those involving chemical reactions. In ab initio MD (AIMD) simulations, the interatomic potentials are accurately determined by solving Schrodinger’s equation \cite{car1985unified}. However, the computational cost of AIMD is not affordable to simulate large systems. Machine learning interatomic potentials (MLIPs) are promising alternatives that can achieve ab initio level accuracy with high efficiency by using machine learning models to fit the ab initio energies and forces \cite{butler_machine_2018, noe_machine_2020, unke_machine_2021}. Among them, Graph Neural Networks (GNNs) are widely used by learning the interatomic potential directly from the atomic coordinates and chemical symbols without relying on any predefined functional form or prior knowledge \cite{gilmer2017neural}. GNNs can also leverage physical systems' inherent symmetries and invariances, such as translation, rotation, and permutation, to improve their performance \cite{schnet, dimenet, spherenet}. Equivariant neural networks have also been developed to preserve the equivariance of molecular systems to improve the performance of interatomic potential predictions \cite{thomas2018tensor, batzner20223, egnn, painn, gasteiger2021gemnet}.

Although MLIPs have been extensively studied, their performance and transferability are still limited by the scarcity of training data due to expensive ab initio calculations. Various self-supervised learning (SSL) pre-training approaches have been explored to learn transferable representations from abundant unlabeled data \cite{graph_infomax, hassani2020contrastive, qiu2020gcc, hu2019strategies}. The pre-trained model is then fine-tuned to extract more task-specific information from the limited labeled data. For example, Wang et al. propose a graph-level SSL method for 2D molecular graphs called MolCLR \cite{contrast}. Uni-Mol \cite{zhou2023uni} is an SE(3) equivariant transformer architecture for pre-training molecular representation, where SSL tasks of masked atom prediction and 3D position recovery are used. However, they focus on the task of molecular property prediction, and only a few works investigate the pre-training methods of interatomic potential. 
In addition to designing SSL tasks, the acquisition of pre-training datasets is also an issue for interatomic potential.
Current pre-training methods for interatomic potential \cite{zhang2022dpa, denoise} rely on existing datasets generated by expensive ab initio calculations such as OC20 \cite{chanussot2021open} (catalytic systems) and ANI-1 \cite{smith2017ani} (small molecules), which are unscalable due to the expensive cost in generating sufficient pre-training data with ab initio calculations and have limited generalization ability to new molecular systems that are not covered by the existing datasets. Gardner et al. \cite{gardner2023synthetic} recently propose a synthetic pre-training method that avoids such problem, but it relies on existing well-trained MLIPs which are not always available.

In this work, we propose an SSL framework by first conducting self-supervised pre-training on the cheap and easy-to-obtain CMD data \cite{kaminski2001evaluation} and then fine-tuning on the labeled data generated by AIMD to learn the task-related information. Under this framework, large-scale pre-training datasets can be easily generated with limited cost for a specific molecular system, enhancing the pre-training ability and avoiding the domain gap between the systems of pre-training data and test data. 
Besides, we also propose a geometry-enhanced self-supervised learning that involves three complementary tasks, including restoring the masked atoms with noisy coordinates, predicting noise with the masked atoms \cite{denoise}, and contrastive learning with 3D network \cite{stark20223d} to better extract topology and 3D geometric information from CMD data. We demonstrate that the CMD structures without labels benefit interatomic potential predictions. To prove the effectiveness of our method, we pre-train MLIPs with GPIP and apply them to various challenging benchmarks for interatomic potential prediction. To better evaluate the capability of MLIPs, we also develop an electrolyte solution dataset that includes many more types of elements and configurations. 
We emphasize that the computational complexity of GPIP is negligible compared to ab initio calculations for obtaining accurate labels for the training set. Although increasing the number of training data can also improve accuracy, the computational cost could be hundreds of times higher than that of GPIP due to ab initio calculations. 

We list the main contributions: (1) We propose a geometric learning framework for MLIPs, GPIP, consisting of two components: geometric structure generation and geometry-enhanced pre-training. (2) We demonstrate that the unlabeled geometric structures produced by CMD simulations benefit interatomic potential predictions with very small computational costs. (3) We design a geometry-enhanced self-supervised pre-training method for molecular structural data. (4) We evaluate our method using various MLIPs and datasets covering various molecular systems. Our method shows consistent and robust performance in all the experiments. 

\section{Results}\label{sec2}

\subsection{Preliminaries}
 
A certain molecule conformation can be uniquely represented by $n$ atoms with atomic numbers $Z=(z_1,...,z_n)$ and atomic positions $R=(r_1,...,r_n)$. The atom features are first initialized using embedding: $X^0 = A_Z$ where $X^0 = (x^0_1,...,x^0_n)$ and $A_Z = (a_{z_1},...,a_{z_n})$. The atom embeddings $\{a_{z_i}\}_{i=1}^n$ are randomly initialized and updated during training. Then, the graph neural network is used to further iteratively update the atom features through the message-passing schemes:

\begin{equation}
    m_{i j}^{l}\,=\,f_{m}^{l}(x_{i}^{l},x_{j}^{l},{\alpha}_{i j})
\end{equation}

\noindent where $\alpha_{ij}$ represents the interaction between atoms $i$ and $j$ (usually, it is the distance or adjacency between atoms $i$ and $j$), $x_{i}^{l}$ and $ x_{j}^{l}$ are features for atoms $i$ and $j$ respectively at the $l^{th}$ layer.  Next, the messages $\{m_{i j}^{l}\}_{j\in N_i}$ at the $l^{th}$ layer from neighbors $\mathcal{N}_i$ of atom $i$ are aggregated and the corresponding atom feature is updated by the function $f_{u}^{l}$:
\begin{equation}
    x_{i}^{l+1}\,=\,f_{u}^{l}(x_{i}^{l},\sum_{j\in \mathcal{N}_{i}}m_{i j}^{l}).
\end{equation}

Finally, the updated features are leveraged in various downstream tasks, such as predicting the potential energies and forces.

In this work, we implement our proposed GPIP on both invariant and equivariant baseline models to prove its effectiveness. The selected baseline models cover a wide range of GNNs and exhibit competitive performance on various benchmark datasets for interatomic potentials. 

\subsection{GPIP framework}

\textbf{Overview of GPIP.}
GPIP is a geometric structure learning framework that leverages low-cost structural data to learn richer configuration information. First, structural data of target molecular systems are generated by CMD simulations with empirical force fields. Next, we pre-train on the CMD structural data to learn both topology and spatial structure information. We note that, for complex molecular systems with periodic boundary conditions (PBCs), a single pre-training task cannot effectively learn structure representations. Therefore, we propose a geometry-enhanced self-supervised learning approach that consists of three complementary pre-training tasks: restoring the masked atoms with noisy coordinates, predicting noise with the masked atoms \cite{denoise}, and contrastive learning with 3D network. GPIP is model-agnostic and can be implemented on different types of GNNs. The overall architecture of GPIP is illustrated in Figure 1.

\noindent \textbf{CMD simulations with empirical force fields.}
Empirical force fields are widely used in CMD simulations because they are computationally efficient and can handle large and complex systems. Despite the limitations such as lower accuracy and transferability compared with ab initio methods, CMD simulations can provide abundant structural information useful to improve the performance of MLIPs. There are force field models that cover most of the periodic table, such as the universal force field (UFF) \cite{rappe1992uff} and the all-atom optimized potentials for liquid simulations (OPLS-AA) force field \cite{kaminski2001evaluation}.

\noindent \textbf{Restore Masked Atoms with Noisy Coordinates.} 
We adopt a masking mechanism to enhance the ability of GNNs to capture the topology information in molecular graphs. By learning the representations of masked atoms via global interpolation of visible atoms, Masked Autoencoders \cite{he2022masked} can capture the topology information that what atoms are close to or connected with each other. 
The masking task process is illustrated in task 1 of Figure 1. We first sample a subset of nodes $\tilde{N}$ and mask their features with a mask token [MASK]. Then we use the GNN encoder to convert the masked conformation into a latent code $\mathbf{H}$ and attempt to restore the masked features of the nodes in $\tilde{N}$ with the decoder, based on the partially observed node signals and atom coordinates. To prevent the trivial solution of GNN encoder caused by atom-type information leakage via 3D coordinates and learn more useful topology information, we add small noises to atom coordinates and leverage the re-mask technique on latent code $\mathbf{H}$ \cite{hou2022graphmae}. We replace the feature vectors of masked node indices in $\mathbf{H}$ with another mask token [DMASK]. 
To avoid bias from predicting a fixed set of masked nodes that may not represent the whole graph structure, we use a uniform sampling strategy to select the nodes to mask. Furthermore, the [MASK] token may lead to a discrepancy between training and inference as it is absent during inference. Therefore, we replace the [MASK] token with a random token with a low probability (e.g., 15\% or less). 
In this task, we use scaled cosine error as the criterion for reconstructing original node features, allowing more accurate reconstructions.

\noindent \textbf{Predict Noise with Masked Atoms.} 
To extract the spatial structure information, we leverage the denoising autoencoders (DAEs) \cite{vincent2008extracting} to pre-train GNNs. By purifying the noisy input, DAE can help to capture atomic position information of the pre-training data. 
As shown in task 2 of Figure 1, we perturb molecular configurations by adding random noises $\mathbf{\mathcal{E}} \in \mathbb{R}^{n \times 3}$ drawn from a Gaussian distribution to the coordinates of atoms. The perturbed coordinates are denoted as $\hat{R}=R+\mathbf{\mathcal{E}}$. We then train an autoencoder $g_{\phi}\circ f_{\theta}$ to minimize the reconstruction error $\|g_{\phi}(f_{\theta}(\tilde{X}^0,\hat{R}))-\mathcal{E}\|^{2}$, where $\tilde{X}^0$ represent the features of masked atoms. The encoder $f_{\theta}$ and decoder $g_{\phi}$ are parameterized by two sets of trainable parameters $\theta$ and $\phi$. The objective of a standard denoising autoencoder is to predict the unperturbed coordinates $R$. However, GNNs with skip connections can easily restore $R$ from $\hat{R}$, which cannot capture useful spatial information. Therefore, we opted to predict the noise $\mathcal{E}$ instead of predicting the unperturbed $R$. To make this task more challenging, we randomly sampled a certain proportion of nodes and masked their features. 
The magnitude of noise added to the coordinates is a crucial hyperparameter in the denoising task. The noise variance $\sigma$ should be large enough to force the GNNs to learn meaningful spatial structural representations, but not so large that causes excessive distribution shift between clean and noisy conformations.

\noindent \textbf{Constrastive Learning with 3D Network.} 
To further extract global 3D information, we employ contrastive learning to enforce GNN to learn the representations of a 3D network. The 3D network encodes the pairwise Euclidean distances of all atoms and therefore completely describes the 3D structures. The pairwise distances of a molecular configuration are first mapped to a higher dimensional space using sine and cosine functions with different frequencies as edge features, while node features are all set to the same learnable vector that initialized with a standard Gaussian distribution \cite{stark20223d}. The features are updated by multiple message-passing layers that work on fully connected graph data to generate the 3D latent code $\mathbf{H}_b$. 
Meanwhile, the same molecular configuration with perturbed coordinates are fed into a GNN $f^{a}$ to generate the GNN latent code $\mathbf{H}_a$. 
Next, we maximize the mutual information between the output features of GNN $f^{a}$ and 3D network $f^{b}$, as shown in task 3 of Figure 1. In this way, GNN $f^{a}$ can learn more 3D information from 3D network $f^{b}$ which can help improve downstream interatomic potential predictions. 
The task 3 can also be seen as a contrastive distillation process, where the student GNN learns from the teacher 3D network to capture 3D information.

\subsection{Experiments}\label{sec3}

\noindent \textbf{MD17: MD Trajectories of small molecules.}
MD17 dataset \cite{chmiela2017machine} consists of small organic molecules generated by AIMD simulations.
The results of GPIP-based models and baseline models are listed in \cref{tab:md17}. It is clear that the performances of GPIP-based SchNet, DimeNet, SphereNet, and GemNet-T are all better than their corresponding baseline models. We also perform ablation studies to study the effect of each task in GPIP and the results are summarized in Supplementary Table S3.

\noindent \textbf{ISO17: MD Trajectories of C$_7$O$_2$H$_{10}$ isomers.}
ISO17 dataset \cite{ramakrishnan2014quantum, schnet} consists of MD trajectories of C$_7$O$_2$H$_{10}$ isomers generated by AIMD simulations. 
We first use a small dataset of 20,000 configurations and a large dataset of 400,000 configurations from ISO17 dataset respectively for fine-tuning. Two different settings are considered here to evaluate the generalization ability of MLIPs across conformational space. For the unknown molecules/unknown conformations task, the molecules in the test set are not in the training set. As shown in Figures 2(a) and Figures 2(b), GPIP-based models significantly outperform their non-pre-trained counterparts on both tasks, even though the fine-tuning dataset is very large. 
To further study the data efficiency, we randomly sampled various amounts of data from the original dataset for fune-tuning while the test sets remain unchanged. The results of force predictions with SchNet and EGNN baseline models are shown in Figures 2(c) and Figures 2(d), respectively. The GPIP-based models perform better than the non-pre-trained counterparts throughout the experiments. We find that using only 10,000 training data, the GPIP-based SchNet achieves an MAE of 9.95, much lower than the non-pre-trained SchNet with an MAE of 18.08. This performance is even better than the non-pre-trained SchNet trained using 10 times more training data, which results in an MAE of 10.55. Therefore, we demonstrate that GPIP-based MLIPs are more data efficient than those trained from scratch in achieving comparable performance. Although the MAE continues to decrease with increasing training data, considering the computational costs for acquiring high-precision labeled training data, GPIP can achieve comparable accuracy with far fewer computational resources. For example, the GPIP-based SchNet achieves an MAE of 6.16 with 100,000 training data, while the non-pre-trained SchNet achieves an MAE of 6.81 with 200,000 training data. Using GPIP, the additional computational costs include about 230 CPU core hours for CMD data generation and 1.2 hours of GPU computation (using one NVIDIA A100 card) for pre-training. In comparison, obtaining an additional 100,000 labeled ISO17 data by ab initio calculation requires about 53,000 CPU core hours.

\noindent \textbf{Liquid Water with PBCs.}
Training MLIPs for extended systems with PBCs is more difficult than for single-molecule systems due to structural complexity. Water is arguably the most important molecular liquid in biological and chemical processes, and its complex thermodynamic and phase behaviors pose great challenges for computational studies \cite{fu2022forces}.
In addition, high accuracy of force alone is not sufficient for an effective MD simulation \cite{fu2022forces}. Therefore, we further evaluate the MLIPs through MD simulations with more quantitative metrics. Specifically, we first train MLIPs and evaluate the accuracy of force. Then we perform MD simulations to further evaluate simulation stability \cite{fu2022forces} and equilibrium statistics including radial distribution function (RDF) and diffusion coefficient. By initializing from 5 randomly sampled testing configurations, we perform 5 simulations of 500 $\text{ps}$ with a time step of 1 $\text{fs}$ at the temperature of 300 K. Then the MD trajectories are analyzed to evaluate the MLIPs. As shown in \cref{tab:water}, GPIP-based models improve not only the accuracy of force prediction but also other key metrics in MD simulations.

\noindent \textbf{Electrolyte Solutions with PBCs.}
Electrolyte solutions involve more types of elements and more complex interatomic interactions compared to liquid water. Since the configurational space expands combinatorially with the increase of element types, training MLIPs for such complex systems is more complicated and attracts more and more attention \cite{unke_machine_2021}. Researches have shown that MLIPs without explicitly considering long-range forces perform well in many cases, e.g., phase diagram of water \cite{phase-water}, lithium transport in the isotropic bulk solid-state electrolytes \cite{staacke2021role}, and chemical reaction in molten alkali carbonate-hydroxide \cite{mondal2022modeling}, but result in poor predictions for systems exhibiting phenomena such as long-range polarization and electrostatics \cite{anstine2023machine}. We construct a dataset of electrolyte solutions and test the performance of GPIP on it. 
\cref{tab:elect} shows that GPIP significantly improves the performances of all tested MLIPs. We also perform ablation studies to test the effect of each task in GPIP and the results are summarized in Supplementary Table S4. 

\section{Discussion}\label{sec3}

The effectiveness of GPIP is attributed to the ability of self-supervised tasks to extract structural information from the data generated by CMD simulations with empirical force fields efficiently. Although not as accurate as ab initio methods, CMD simulations can provide physically reasonable atomic structures as the empirical force fields are based on physical formulas. Moreover, when sampled from long-time CMD simulations at different temperatures and pressures, abundant structural data can be sampled from a large space, covering the fine-tuning datasets in downstream tasks. This coverage can be visualized through the UMAP dimensionality reduction method \cite{mcinnes2020umap}, as shown in Figure 3. For all four datasets used in this work, the CMD datasets adequately cover the downstream task data and are spread over a larger space. Therefore, the structural information learned from CMD data benefits downstream tasks. The CMD data can be used also for other SSL methods. Supplementary Table S5 shows that CMD data works well with the pre-training method of Uni-Mol \cite{zhou2023uni} for interatomic potential prediction. 

Another significant benefit of using CMD-generated data for pre-training is the reduction of computational cost. Compared to ab initio calculations, implementing CMD with empirical force fields requires significantly fewer core hours to calculate potential energies and forces. CPU parallelization and GPU acceleration add further advantages to this reduction. As indicated in Supplementary Table S2, datasets consisting of millions of atoms can be generated within hours using commonly used MD software packages such as LAMMPS, whereas the increases in accuracy of predicted labels are notable. GPIP provides future practitioners with a means to enhance MLIPs’ performance with relatively accessible computing power requirements.

In conclusion, we propose a geometric learning framework for MLIPs called GPIP, which consists of two components: geometric structure generation and geometry-enhanced pre-training. Plenty of 3D geometric structures of various molecular systems can be generated efficiently by performing CMD simulations with empirical force fields. We then design a self-supervised pre-training scheme incorporating three tasks: masking, denoising, and contrastive learning. These tasks aim to capture both topology and spatial structure information from the unlabeled geometric structures produced by CMD simulations. We show that the three tasks are complementary and a single pre-training task may not improve the accuracy significantly. We evaluate our method using various MLIPs and datasets that cover a wide range of molecular systems, from simple molecules to complex systems with PBCs. In addition to the most commonly used MD17, ISO17, and liquid water datasets, we present an electrolyte solution dataset that includes more types of elements and more complex configurations for benchmarking. Our experimental results show that the GPIP-based models consistently and significantly outperform the corresponding baseline models, demonstrating remarkable effectiveness and robustness. The computational costs of GPIP are negligible compared to ab initio calculations to obtain more labeled training data. This work is highly applicable for performing MD simulations using MLIPs, especially for complex molecular systems where sufficient training data is hard to obtain.

\backmatter

\section{Methods}\label{sec4}

\subsection{Dataset}

We evaluate the performance of GPIP on multiple benchmark datasets ranging from simple systems including MD17 \cite{chmiela2017machine} and ISO17 \cite{ramakrishnan2014quantum}) to complex systems with PBCs including liquid water \cite{fu2022forces} and our dataset of electrolyte solutions. 

\noindent \textbf{Pre-training Dataset.}

To obtain sufficient configurations that can cover the test set, we sample a large number of configurations from CMD trajectories to serve as the pre-training datasets. 
All CMD simulations are conducted with LAMMPS software package\cite{thompson2022lammps} and the timestep is set to 1 $\text{fs}$. 

For MD17, the pre-training dataset is composed of 160,000 configurations sampled from CMD trajectories performed on eight different organic molecules; and for ISO17, the CMD simulations are performed on 100 isomers with atoms C$_7$O$_2$H$_{10}$, which generates 2,000,000 configurations. These configurations are sampled with time resolution of 1 $\text{ps}$ in vacuum at different temperatures varying from 300 K to 1000 K with intervals of 100 K under NVT ensembles, and the force field parameters were taken from OPLS-AA force field \cite{kaminski2001evaluation}. 

The pre-training dataset for liquid water with PBCs is generated with TIP3P force field parameters \cite{jorgensen1983comparison}. The trajectories contain 64 water molecules and are calculated at 10 different temperatures from 100 K to 1000 K with intervals of 100 K under NVT ensembles and 100,000 configurations are sampled with a resolution of 1 $\text{ps}$. The density of the configurations is determined by AIMD calculation at 1 atm and 300 K with an NPT ensemble. 

As for the electrolyte solution pre-training dataset, we select different combinations of ions and solvents to be the initial configurations, which include $[\rm Na]^{+}$ or $[\rm Li]^{+}$ cations, $[\rm PF_{6}]^{-}$ or  $[\rm Tf_{2}N]^{-}$ anions, and DME or 1:1 EC+DMC solvent. The concentrations of electrolyte solutions are set to be 1 M or 4 M. The force field parameters were taken from OPLS-AA force field \cite{kaminski2001evaluation}. The CMD trajectories on 16 different electrolytes are calculated at temperatures from 100 K to 1000 K with intervals of 100 K. Each contains 24 (1 M) or 48 (4 M) pairs of ions and 192 (1 M) or 72 (4 M) solvent molecules. The size of the periodic box of each electrolyte solution is determined by a 5 $\text{ns}$ simulation with NPT ensemble at 300 K, 1 atm, and an equilibrium stage that anneals the electrolyte solution from 1000 K to 300 K in 10 $\text{ns}$ was performed. Eventually, 160,000 configurations of electrolyte solution are sampled with NVT ensemble at different temperatures for 100 $\text{ns}$ with a time resolution of 1 $\text{ps}$.

\noindent \textbf{Downstream Dataset.}

The MD17 \cite{chmiela2017machine}, ISO17 \cite{ramakrishnan2014quantum,schnet}, and liquid water \cite{fu2022forces} datasets used for fine-tuning are taken from previous studies. The structures of electrolyte solutions are sampled from 16 CMD trajectories with different compounds and concentrations whereas the atomic forces and total energies were determined by density functional theory (DFT) calculations using PBE exchange-correlation functional\cite{perdew1996generalized} and with the Projector Augmented-Wave (PAW) pseudo-potential\cite{blochl1994projector}. A single gamma k-point is used to sample the Brillouin zone and the cutoff energy for the plane-wave-basis set is set to be 400 eV and the electronic self-consistency is considered to be achieved if the change of total energy between two steps is smaller than $10^{-5}$ eV. Each configuration contains 3 (1 M) or 6 (4 M) pairs of ions and 24 (1 M) or 9 (4 M) solvent molecules. The equilibrium and sampling stages of the fine-tuning data set are identical to the pre-training dataset, and the DFT calculation was performed on randomly selected configurations using the VASP code\cite{blochl1994projector} with a plane-wave basis set. There are in total 8000 configurations sampled and calculated.

\subsection{Pre-training Settings}
We apply three SSL tasks (masking, denoising, and contrastive learning) to learn the spatial and topological information from pre-training data. The final loss function is defined as follows:
\begin{equation}
    {\mathcal{L}}_{\text{final}} = {\mathcal{L}}_{\text{mask}} + \mathcal{\alpha}{\mathcal{L}}_{\text{denoise}} + \mathcal{\beta}{\mathcal{L}}_{\text{contrast}} ,
    \label{alpha}
\end{equation}
where $\alpha$ and $\beta$ are hyperparameters for balancing the losses of three tasks,
${\mathcal{L}}_{\text{mask}}$, ${\mathcal{L}}_{\text{denoise}}$, and ${\mathcal{L}}_{\text{contrast}}$ are the corresponding losses for masking, denoising, and contrastive learning. 

In task 1, we utilize the cosine error as the criterion for reconstructing original node features. Additionally, we downweight the easy samples in the training set by scaling the cosine error with a power of $\gamma$ where $\gamma \geq 1$.

\begin{equation}
{\mathcal{L}}_{\text{mask}}={\frac{1}{\left|{\tilde{N}}\right|}} \sum_{i\in{\mathcal{C}}}\left(1-{\frac{{x_{i}^0}^{T}x_{i}^\text{out}}{\|x_{i}^0\|\cdot\|x_{i}^\text{out}\|}}\right)^{\gamma},{\gamma}\geq1,
\end{equation}
where $x_i^{0}$ denotes original feature, $x_i^{out}$ is the output of the decoder, $\mathcal{C}$ is the set of indices of the masked nodes $\tilde{N}$, and the scaling factor $\gamma$ is a hyperparameter that can be adjusted according to different datasets.

In task 2, the GNN model is trained to predict the additional noise under the challenge of randomly masked atoms, and the objective function is
\begin{equation}
    {\mathcal{L}}_{\text{denoise}} = \mathbb{E}_{p(\hat{R},\tilde{X})}\left[\left\|f_{\theta}(\tilde{X}, \hat{R})-\mathcal{E}\right\|_2^{2}\right],
\end{equation}
where $f_{\theta}$ denotes an invariant/equivariant GNN model parameterized by $\theta$ and $p(\hat{R},\tilde{X})$ measures the probability distribution of perturbed molecular conformations.

In task 3, the GNN is enforced to learn the 3D information from 3D network by maximizing the mutual information between the latent features of these two nets. The corresponding loss function is:
\begin{equation}
{\mathcal{L}}_{\text{contrast}}=-\frac{1}{n} \sum_{i=1}^{n}\left[\log \frac{e^{\operatorname{sim}\left(h_{i}^{a}, h_{i}^{b}\right) / \tau}}{\sum_{\substack{k=1 \\ k \neq i}}^{N} e^{\operatorname{sim}\left(h_{i}^{a}, h_{k}^{b}\right) / \tau}}\right],
\end{equation}
where $\operatorname{sim}\left(h^{a}, h^{b}\right)=h^{a} \cdot h^{b} /\left(\left\|h^{a}\right\|\left\|h^{b}\right\|\right)$ is the cosine similarity, $h^{a}$ and $h^{b}$ are latent codes of GNN and 3D network, respectively, and $\tau$ is a hyperparameter that can be seen as the weight for the most similar negative pair.

The pre-training settings are determined according to the dataset during the experiments. For the MD17 dataset, all invariant or equivariant GNN models are trained for 3 epochs with an initial learning rate of 0.001 and a weight decay of 0.01 during pre-training. We set $\alpha=1$ and $\beta=1$ in \cref{alpha}. The models are first pre-trained on the CMD data of all molecules and then fine-tined on the labeled data of each molecule separately. 
We utilize the Adam optimizer with a batch size of 100. As for the ISO17 dataset, we train the GNN models for 3 epochs with GPIP. The mask rate and batch size are set to 0.15 and 100, respectively. We set $\alpha=1$ and $\beta=0.5$ in \cref{alpha}. The Adam optimizer is utilized with a learning rate of $5\times 10^{-5}$ and a weight decay of 0.01.

For the two datasets with PBCs, we train the GNN models 
for 3 epochs with GPIP. We employed the Adam optimizer with an initial learning rate of $4\times10^{-3}$ and a weight decay of 0.01. Additionally, we set the pre-training loss with $\alpha=1$ and $\beta=0.1$ in \cref{alpha}. 
We use a mask rate of 0.15 and a batch size of 100 for the liquid water dataset, and a mask rate of 0.05 and a batch size of 3 for the electrolyte solution dataset.

\subsection{Fine-tuning Settings}
All experiments are implemented by PyTorch 1.8.0. We construct the loss function to fit the total energy $E$ as well as the force $F_i$ on each atom:
\begin{equation}
    \mathcal{L}_{\text{fine-tune}} = ||E-\hat{E}||^{2}+\frac{100}{n}\sum_{i=0}^{n}\left|\left|\mathbf{F}_{i}-\left(-\frac{\partial\hat{E}}{\partial\mathbf{r}_{i}}\right)\right|\right|^{2},
\end{equation}
where $n$ is the number of atoms in a data sample, $\mathbf{r}_{i}$ represents the coordinate of atom $i$, and $\hat{E}$ denotes the predicted energy. The training setup varies when GNNs are trained on different datasets, and we show the details as follows. 

\textit{MD17.} The models are trained for 500 epochs. Both training and validation sets consist of 1,000 configurations. The parameters of the pre-trained embedding layers and message-passing layers are transferred during the fine-tuning process. All hyperparameters for each GNN are based on the recommendations from the previous literature \cite{spherenet}. More details regarding the GNNs implemented in this work can be found in Supplementary Table S6.

\textit{ISO17.} We utilize two training sets during fine-tuning: a small dataset of 20,000 configurations and a large dataset with 400,000 configurations. The small dataset is randomly sampled from the large training dataset. For validation and test datasets, we follow the splitting strategy reported in the previous literature for ISO17\cite{schnet}. This makes it convenient for us to evaluate the effectiveness of GPIP. We train the GNN with AdamW \cite{adamw} optimizer, and the hyperparameters are provided in Supplementary Table S7.

\textit{Liquid Water with PBCs.} We benchmark different models with training datasets with the sizes of 950 for DimeNet and GemNet-T and 9,500 for SchNet and use the remaining 10,000 structures for testing. We perform 5 simulations of 500 ps by initializing from 5 randomly sampled testing configurations, with a time step of 1 fs, at the temperature of 300 K. The other experimental settings are the same as \cite{fu2022forces}.

\textit{Electrolyte solutions with PBCs.} The training, validation, and test sets contain 240, 60, and 500 configurations. For training hyperparameters, we use the Adam with a learning rate of $5\times10^{-4}$ and a weight decay of 0.8 and train the model for 800 epochs. More details can be seen in Supplementary Table S8.

\section*{Data availability}
The data used for pre-training and downstream tasks are available in the figshare database \cite{data}. 

\section*{Code availability}
The source code of GPIP framework is available at GitHub: \url{https://github.com/cuitaoyong/GPIP} \cite{code}. 

\section*{Acknowledgments}
This work was supported by the National Key R\&D Program of China (NO.2022ZD0160101).
M.S. was partially supported by Shanghai Committee of Science and Technology, China (Grant No. 23QD1400900).

\section*{Author contributions}
M.S. and S.Z. conceived the idea and led the research. T.C. developed the codes and trained the models. C.T. generated datasets and performed experiments and analyses. Y.L. and X.G. contributed technical ideas for datasets and experiments. L.B., Y.D., and W.O. contributed technical ideas for self-supervised methods. T.C., C.T., M.S., and S.Z. wrote the paper. All authors discussed the results and reviewed the manuscript. 

\section*{Competing interests}
The authors declare no competing interests.

\clearpage

\section*{Tables}

\begin{table*}[h]
  \caption{
    \textbf{Results on MD17 dataset.} Force MAEs are reported in unit of [kcal/mol/\AA]. The weights of force over energy in loss functions are set to 100 to in line with the original papers of the baseline models for fair comparisons. Standard deviations are calculated from 5 independent experiments. Results of all baseline models are directly taken or adapted (if the unit varies) from the original papers and standard deviations are not provided.}
  \centering
  \label{tab:md17}
  \resizebox{\textwidth}{!}{
    \begin{tabular}{@{}llcccccccc  @{}}
      \toprule
       &   & SchNet \cite{schnet}  & SchNet-G & DimeNet \cite{dimenet}  & DimeNet-G  &   SphereNet \cite{spherenet}     &  SphereNet-G&GemNet-T \cite{gasteiger2021gemnet}&GemNet-T-G \\  

      \midrule
      \multirow{1}{*}{Aspirin}
       & F                                       & 1.35 & \textbf{0.853}$\pm$\textbf{0.041} &0.499&\textbf{0.309}$\pm$\textbf{0.042}  & 0.430 & \textbf{0.301}$\pm$\textbf{0.032} &0.227& \textbf{0.160}$\pm$\textbf{0.021}  \\

    \midrule
      \multirow{1}{*}{Benzene}
       & F                                      & 0.31 & \textbf{0.271}$\pm$\textbf{0.023} &0.187& \textbf{0.169}$\pm$\textbf{0.008} & 0.178 & \textbf{0.155}$\pm$\textbf{0.035} &0.265&\textbf{0.100}$\pm$\textbf{0.025}  \\
    \midrule
        \multirow{1}{*}{Ethanol}
       & F                                      & 0.39 & \textbf{0.271}$\pm$\textbf{0.022}  &0.230& \textbf{0.146}$\pm$\textbf{0.013}& 0.208 & \textbf{0.150}$\pm$\textbf{0.018}  &0.155&\textbf{0.109}$\pm$\textbf{0.017}    \\

    \midrule
      \multirow{1}{*}{Malonaldehyde}
       & F                                      & 0.66 & \textbf{0.533}$\pm$\textbf{0.022}  &0.383&\textbf{0.264}$\pm$\textbf{0.015} & 0.340 & \textbf{0.261}$\pm$\textbf{0.042}   &0.281&\textbf{0.163}$\pm$\textbf{0.023}  \\

    \midrule
      \multirow{1}{*}{Naphthalene}
       & F                                       & 0.58 & \textbf{0.450}$\pm$\textbf{0.021}    &0.215&\textbf{0.184}$\pm$\textbf{0.010}& 0.178 & \textbf{0.093}$\pm$\textbf{0.021} &0.101&\textbf{0.074}$\pm$\textbf{0.012}   \\

    \midrule
      \multirow{1}{*}{Salicylic acid}
       & F                                       & 0.85 &\textbf{0.584}$\pm$\textbf{0.024}    &0.374&\textbf{0.214}$\pm$\textbf{0.016}& 0.360 & \textbf{0.204}$\pm$\textbf{0.040}  &0.231&\textbf{0.153}$\pm$\textbf{0.023} \\
    \midrule
      \multirow{1}{*}{Tolunene}
       & F                                      & 0.57 & \textbf{0.412}$\pm$\textbf{0.013}  &0.216& \textbf{0.090}$\pm$\textbf{0.023}& 0.155 & \textbf{0.090}$\pm$\textbf{0.009}  &0.109&\textbf{0.081}$\pm$\textbf{0.011} \\

    \midrule
      \multirow{1}{*}{Uracil}
       & F                                       & 0.56 & \textbf{0.481}$\pm$\textbf{0.024}     &0.301&\textbf{0.261}$\pm$\textbf{0.019}& 0.267 & \textbf{0.195}$\pm$\textbf{0.031}  &0.176&\textbf{0.123}$\pm$\textbf{0.019} \\

      \bottomrule
      \multicolumn{7}{l}{The results of GPIP-based models are shown in bold.}
    \end{tabular}
}
\end{table*}

\begin{table*}[h]
\centering \label{tab:water}
\caption{\textbf{Results on Water dataset.} Force MAEs are reported in unit of [meV/\AA], stabilities are reported in unit of [ps], and RDF MAEs are unitless. Diffusivity is computed by averaging 5 runs from 5 random initial configurations and its MAEs are reported in unit of [$10^{-9}\  \mathrm{m}^2/\mathrm{s}$]. Standard deviations are calculated from 5 independent experiments. The force and diffusivity MAEs of baseline models are taken from \cite{fu2022forces} where standard deviations are not provided. The stable trajectories of DimeNet and GemNet-T are too short to calculate the diffusivity.
}
\label{tab:water}
\begin{adjustbox}{width=0.9\textwidth}
\begin{tabular}{lllllll}
\toprule
{} &                  SchNet &                             SchNet-G &      DimeNet                   &      DimeNet-G&                      GemNet-T  &                                                      GemNet-T-G \\
\midrule
Force              &                    $9.5$  &               $\textbf{8.31}$$\pm$$\textbf{0.17}$  &     3.5        &   $\textbf{2.73}$$\pm$$\textbf{0.08}$  &        $5.0$  &               $\textbf{2.91}$$\pm$$\textbf{0.16}$   \\

Stability         &        $232$$\pm$$59$  &      $\textbf{235}$$\pm$$\textbf{30}$  &    $4$$\pm$$4$      &  $\textbf{18}$$\pm$$\textbf{4}$  &   $6$$\pm$$7$  &            $\textbf{25}$$\pm$$\textbf{3}$   \\

$\mathrm{RDF}_{(O,O)}$ &          $0.63$$\pm$$0.004$  &  $\textbf{0.45}$$\pm$$\textbf{0.03}$  & $0.46$$\pm$$0.22$  &$\textbf{0.16}$$\pm$$\textbf{0.18}$ &  $0.62$$\pm$$0.48$  &  $\textbf{0.42}$$\pm$$\textbf{0.22}$    \\
$\mathrm{RDF}_{(H,H)}$ &          $0.30$$\pm$$0.02$  &  $\textbf{0.20}$$\pm$$\textbf{0.02}$  &$0.33$$\pm$$0.15$  & $\textbf{0.12}$$\pm$$\textbf{0.14}$ &   $0.35$$\pm$$0.21$  &      $\textbf{0.25}$$\pm$$\textbf{0.25}$    \\
$\mathrm{RDF}_{(H,O)}$ &      $0.57$$\pm$$0.04$  &     $\textbf{0.45}$$\pm$$\textbf{0.04}$  & $0.43$$\pm$$0.17$  &$\textbf{0.19}$$\pm$$\textbf{0.15}$ &     $0.71$$\pm$$0.65$  &   $\textbf{0.55}$$\pm$$\textbf{0.27}$    \\
Diffusivity                     &           $1.90$  &                $\textbf{1.68}$$\pm$$\textbf{0.03}$  &  - &  -  &         -         &                -    \\

\bottomrule
\multicolumn{7}{l}{The results of GPIP-based models are shown in bold.}
\end{tabular}
\end{adjustbox}
\end{table*}

\begin{table*}[h]
  \caption{
    \textbf{Results on Electrolyte dataset}. 
    Energy and force MAEs are reported in units of [eV] and [eV/\AA], respectively. Standard deviations are calculated from 5 independent experiments.
    }
  
  \centering
  \label{tab:elect}
  \resizebox{\textwidth}{!}{
    \begin{tabular}{@{}lcccccc  @{}}
      \toprule
       &   SchNet & SchNet-G & PaiNN \cite{painn}                       & PaiNN-G & EGNN \cite{egnn} &  EGNN-G  \\            
      
      \midrule
     Energy & $0.713$ $\pm$$0.085  $                               & $\textbf{0.516}$$\pm$$\textbf{0.072}  $                     & $0.397$$\pm$$0.054   $                                     & $\textbf{0.163} $$\pm$$\textbf{0.022}  $          & $1.006$$\pm$$0.056$ &   $\textbf{0.468}$$\pm$$\textbf{0.013}$                   \\
     Force & $0.161 $$\pm$$0.024$& $\textbf{0.135}$$\pm$$\textbf{0.022}$                &$ 0.090$$\pm$$0.017$     &                              $\textbf{0.061}$$\pm$$\textbf{0.008}$      &     $0.343$$\pm$$0.044$  &$\textbf{0.144}$$\pm$$\textbf{0.016}$              \\

      \bottomrule
      \multicolumn{7}{l}{The results of GPIP-based models are shown in bold.}
    \end{tabular}
     }
\end{table*}

\clearpage

\section*{Figure Legends/Captions}

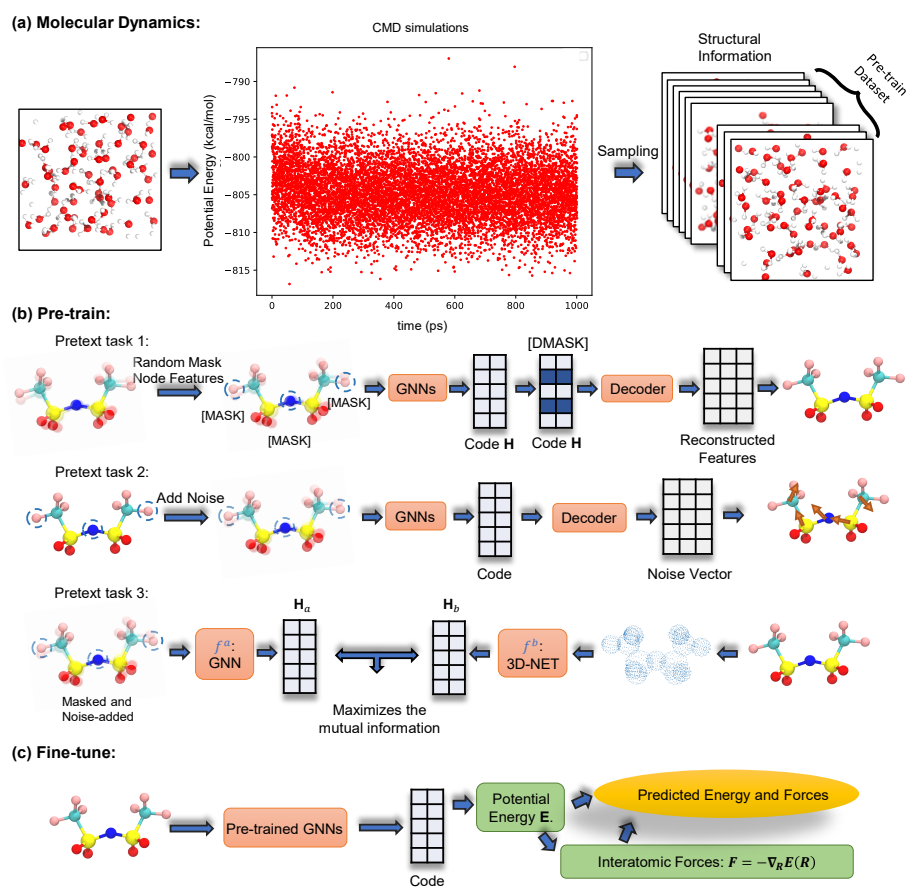
\textbf{Figure 1.} Overview. (a) CMD simulations are performed to generate datasets for pre-training. The structural data are sampled from simulation trajectories with different potential energies at different simulation times. (b) The pre-training method contains three complementary tasks: masked atom restoration with noisy coordinates, noise prediction with masked atoms, and contrastive learning with 3D Nets. (c) GNNs are fine-tuned using labeled data after pre-training.

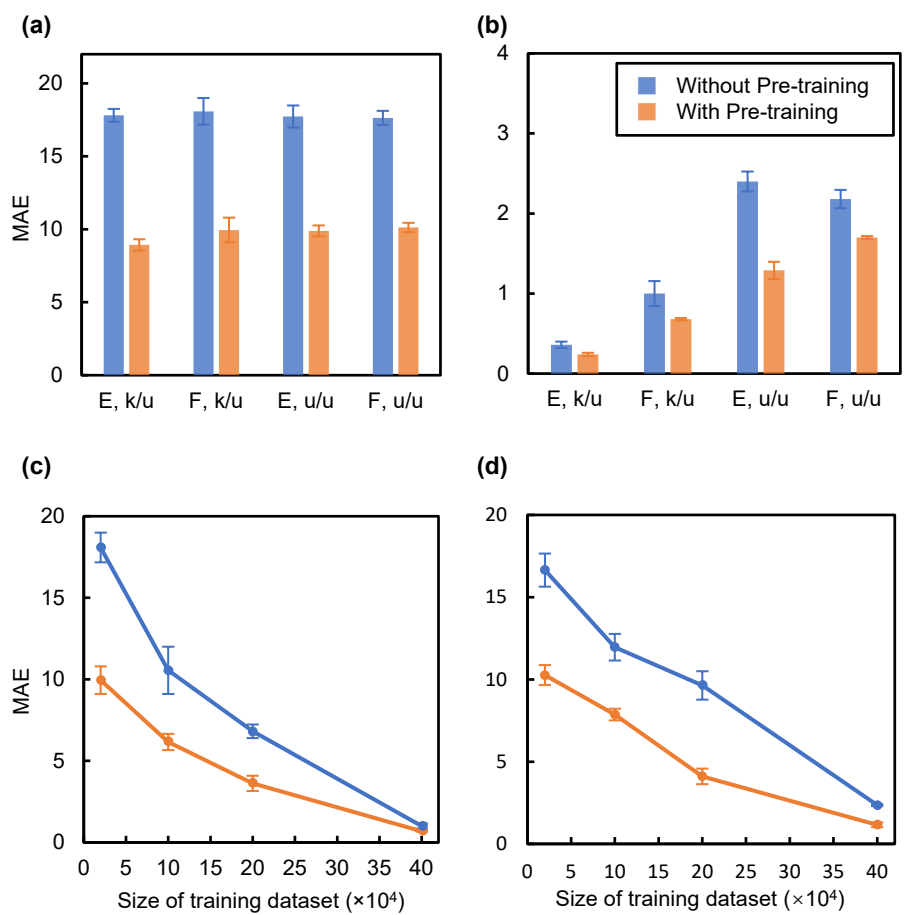
\textbf{Figure 2.} Results on ISO17 dataset. (a) Energy and force MAEs using 20,000 training data with SchNet baseline. (b) Energy and force MAEs using 400,000 training data with SchNet baseline. (c) Force MAEs with training data ranging from 20,000 to 400,000 with SchNet baseline. (d) Force MAEs with training data ranging from 20,000 to 400,000 with EGNN baseline. k/u: train on known molecules and test on unknown conformations. u/u: train on unknown molecules and test on unknown conformations. Energy and force MAEs are reported in units of [kcal/mol] and [kcal/mol /\AA], respectively. Error bar denotes the standard deviation from 5 independent experiments. The error bars of results for 400,000 training data in (c) and (d) are smaller than the data points. 

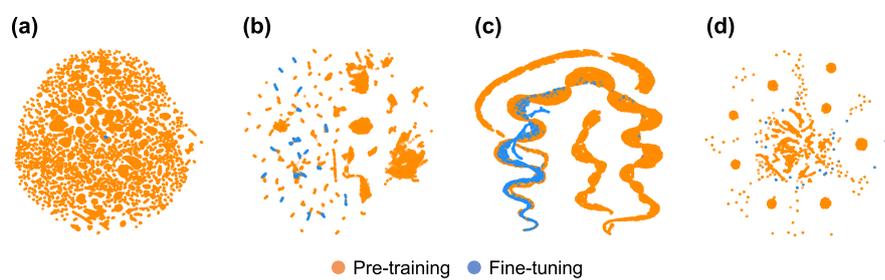
\textbf{Figure 3.} Projection of the pre-training and fine-tuning data onto the embedding of the SchNet-GPIP model using UMAP \cite{mcinnes2020umap} to perform dimensionality reduction. (a) MD17. (b) ISO17. (c) Water. (d) Electrolyte.



\begin{thebibliography}{42}
\ifx \bisbn   \undefined \def \bisbn  #1{ISBN #1}\fi
\ifx \binits  \undefined \def \binits#1{#1}\fi
\ifx \bauthor  \undefined \def \bauthor#1{#1}\fi
\ifx \batitle  \undefined \def \batitle#1{#1}\fi
\ifx \bjtitle  \undefined \def \bjtitle#1{#1}\fi
\ifx \bvolume  \undefined \def \bvolume#1{\textbf{#1}}\fi
\ifx \byear  \undefined \def \byear#1{#1}\fi
\ifx \bissue  \undefined \def \bissue#1{#1}\fi
\ifx \bfpage  \undefined \def \bfpage#1{#1}\fi
\ifx \blpage  \undefined \def \blpage #1{#1}\fi
\ifx \burl  \undefined \def \burl#1{\textsf{#1}}\fi
\ifx \doiurl  \undefined \def \doiurl#1{\url{https://doi.org/#1}}\fi
\ifx \betal  \undefined \def \betal{\textit{et al.}}\fi
\ifx \binstitute  \undefined \def \binstitute#1{#1}\fi
\ifx \binstitutionaled  \undefined \def \binstitutionaled#1{#1}\fi
\ifx \bctitle  \undefined \def \bctitle#1{#1}\fi
\ifx \beditor  \undefined \def \beditor#1{#1}\fi
\ifx \bpublisher  \undefined \def \bpublisher#1{#1}\fi
\ifx \bbtitle  \undefined \def \bbtitle#1{#1}\fi
\ifx \bedition  \undefined \def \bedition#1{#1}\fi
\ifx \bseriesno  \undefined \def \bseriesno#1{#1}\fi
\ifx \blocation  \undefined \def \blocation#1{#1}\fi
\ifx \bsertitle  \undefined \def \bsertitle#1{#1}\fi
\ifx \bsnm \undefined \def \bsnm#1{#1}\fi
\ifx \bsuffix \undefined \def \bsuffix#1{#1}\fi
\ifx \bparticle \undefined \def \bparticle#1{#1}\fi
\ifx \barticle \undefined \def \barticle#1{#1}\fi
\bibcommenthead
\ifx \bconfdate \undefined \def \bconfdate #1{#1}\fi
\ifx \botherref \undefined \def \botherref #1{#1}\fi
\ifx \url \undefined \def \url#1{\textsf{#1}}\fi
\ifx \bchapter \undefined \def \bchapter#1{#1}\fi
\ifx \bbook \undefined \def \bbook#1{#1}\fi
\ifx \bcomment \undefined \def \bcomment#1{#1}\fi
\ifx \oauthor \undefined \def \oauthor#1{#1}\fi
\ifx \citeauthoryear \undefined \def \citeauthoryear#1{#1}\fi
\ifx \endbibitem  \undefined \def \endbibitem {}\fi
\ifx \bconflocation  \undefined \def \bconflocation#1{#1}\fi
\ifx \arxivurl  \undefined \def \arxivurl#1{\textsf{#1}}\fi
\csname PreBibitemsHook\endcsname

\bibitem[\protect\citeauthoryear{Hospital et~al.}{2015}]{hospital2015molecular}
\begin{botherref}
\oauthor{\bsnm{Hospital}, \binits{A.}},
\oauthor{\bsnm{Go{\~n}i}, \binits{J.R.}},
\oauthor{\bsnm{Orozco}, \binits{M.}},
\oauthor{\bsnm{Gelp{\'\i}}, \binits{J.L.}}:
Molecular dynamics simulations: advances and applications.
Advances and applications in bioinformatics and chemistry,
37--47
(2015)
\end{botherref}
\endbibitem

\bibitem[\protect\citeauthoryear{Senftle et~al.}{2016}]{senftle2016reaxff}
\begin{barticle}
\bauthor{\bsnm{Senftle}, \binits{T.P.}},
\bauthor{\bsnm{Hong}, \binits{S.}},
\bauthor{\bsnm{Islam}, \binits{M.M.}},
\bauthor{\bsnm{Kylasa}, \binits{S.B.}},
\bauthor{\bsnm{Zheng}, \binits{Y.}},
\bauthor{\bsnm{Shin}, \binits{Y.K.}},
\bauthor{\bsnm{Junkermeier}, \binits{C.}},
\bauthor{\bsnm{Engel-Herbert}, \binits{R.}},
\bauthor{\bsnm{Janik}, \binits{M.J.}},
\bauthor{\bsnm{Aktulga}, \binits{H.M.}}, \betal:
\batitle{The reaxff reactive force-field: development, applications and future
  directions}.
\bjtitle{npj Computational Materials}
\bvolume{2}(\bissue{1}),
\bfpage{1}--\blpage{14}
(\byear{2016})
\end{barticle}
\endbibitem

\bibitem[\protect\citeauthoryear{Karplus and
  Petsko}{1990}]{karplus1990molecular}
\begin{barticle}
\bauthor{\bsnm{Karplus}, \binits{M.}},
\bauthor{\bsnm{Petsko}, \binits{G.A.}}:
\batitle{Molecular dynamics simulations in biology}.
\bjtitle{Nature}
\bvolume{347},
\bfpage{631}--\blpage{639}
(\byear{1990})
\end{barticle}
\endbibitem

\bibitem[\protect\citeauthoryear{Yao et~al.}{}]{yao_applying_2022}
\begin{botherref}
\oauthor{\bsnm{Yao}, \binits{N.}},
\oauthor{\bsnm{Chen}, \binits{X.}},
\oauthor{\bsnm{Fu}, \binits{Z.-H.}},
\oauthor{\bsnm{Zhang}, \binits{Q.}}:
Applying classical, \textit{Ab Initio} , and machine-learning molecular
  dynamics simulations to the liquid electrolyte for rechargeable batteries
\textbf{122}(12),
10970--11021
\doiurl{10.1021/acs.chemrev.1c00904} .
Accessed 2023-01-17
\end{botherref}
\endbibitem

\bibitem[\protect\citeauthoryear{Kaminski
  et~al.}{2001}]{kaminski2001evaluation}
\begin{barticle}
\bauthor{\bsnm{Kaminski}, \binits{G.A.}},
\bauthor{\bsnm{Friesner}, \binits{R.A.}},
\bauthor{\bsnm{Tirado-Rives}, \binits{J.}},
\bauthor{\bsnm{Jorgensen}, \binits{W.L.}}:
\batitle{Evaluation and reparametrization of the opls-aa force field for
  proteins via comparison with accurate quantum chemical calculations on
  peptides}.
\bjtitle{The Journal of Physical Chemistry B}
\bvolume{105}(\bissue{28}),
\bfpage{6474}--\blpage{6487}
(\byear{2001})
\end{barticle}
\endbibitem

\bibitem[\protect\citeauthoryear{Car and Parrinello}{1985}]{car1985unified}
\begin{barticle}
\bauthor{\bsnm{Car}, \binits{R.}},
\bauthor{\bsnm{Parrinello}, \binits{M.}}:
\batitle{Unified approach for molecular dynamics and density-functional
  theory}.
\bjtitle{Physical review letters}
\bvolume{55}(\bissue{22}),
\bfpage{2471}
(\byear{1985})
\end{barticle}
\endbibitem

\bibitem[\protect\citeauthoryear{Butler et~al.}{}]{butler_machine_2018}
\begin{botherref}
\oauthor{\bsnm{Butler}, \binits{K.T.}},
\oauthor{\bsnm{Davies}, \binits{D.W.}},
\oauthor{\bsnm{Cartwright}, \binits{H.}},
\oauthor{\bsnm{Isayev}, \binits{O.}},
\oauthor{\bsnm{Walsh}, \binits{A.}}:
Machine learning for molecular and materials science
\textbf{559}(7715),
547--555
\doiurl{10.1038/s41586-018-0337-2} .
Accessed 2023-02-11
\end{botherref}
\endbibitem

\bibitem[\protect\citeauthoryear{Noé et~al.}{}]{noe_machine_2020}
\begin{botherref}
\oauthor{\bsnm{Noé}, \binits{F.}},
\oauthor{\bsnm{Tkatchenko}, \binits{A.}},
\oauthor{\bsnm{Müller}, \binits{K.-R.}},
\oauthor{\bsnm{Clementi}, \binits{C.}}:
Machine learning for molecular simulation
\textbf{71}(1),
361--390
\doiurl{10.1146/annurev-physchem-042018-052331} .
Accessed 2023-01-17
\end{botherref}
\endbibitem

\bibitem[\protect\citeauthoryear{Unke et~al.}{}]{unke_machine_2021}
\begin{botherref}
\oauthor{\bsnm{Unke}, \binits{O.T.}},
\oauthor{\bsnm{Chmiela}, \binits{S.}},
\oauthor{\bsnm{Sauceda}, \binits{H.E.}},
\oauthor{\bsnm{Gastegger}, \binits{M.}},
\oauthor{\bsnm{Poltavsky}, \binits{I.}},
\oauthor{\bsnm{Schütt}, \binits{K.T.}},
\oauthor{\bsnm{Tkatchenko}, \binits{A.}},
\oauthor{\bsnm{Müller}, \binits{K.-R.}}:
Machine learning force fields
\textbf{121}(16),
10142--10186
\doiurl{10.1021/acs.chemrev.0c01111} .
Accessed 2023-01-17
\end{botherref}
\endbibitem

\bibitem[\protect\citeauthoryear{Gilmer et~al.}{2017}]{gilmer2017neural}
\begin{bchapter}
\bauthor{\bsnm{Gilmer}, \binits{J.}},
\bauthor{\bsnm{Schoenholz}, \binits{S.S.}},
\bauthor{\bsnm{Riley}, \binits{P.F.}},
\bauthor{\bsnm{Vinyals}, \binits{O.}},
\bauthor{\bsnm{Dahl}, \binits{G.E.}}:
\bctitle{Neural message passing for quantum chemistry}.
In: \bbtitle{International Conference on Machine Learning},
pp. \bfpage{1263}--\blpage{1272}
(\byear{2017}).
\bcomment{PMLR}
\end{bchapter}
\endbibitem

\bibitem[\protect\citeauthoryear{Sch{\"u}tt et~al.}{2017}]{schnet}
\begin{botherref}
\oauthor{\bsnm{Sch{\"u}tt}, \binits{K.}},
\oauthor{\bsnm{Kindermans}, \binits{P.-J.}},
\oauthor{\bsnm{Sauceda~Felix}, \binits{H.E.}},
\oauthor{\bsnm{Chmiela}, \binits{S.}},
\oauthor{\bsnm{Tkatchenko}, \binits{A.}},
\oauthor{\bsnm{M{\"u}ller}, \binits{K.-R.}}:
Schnet: A continuous-filter convolutional neural network for modeling quantum
  interactions.
Advances in neural information processing systems
\textbf{30}
(2017)
\end{botherref}
\endbibitem

\bibitem[\protect\citeauthoryear{Gasteiger et~al.}{2020}]{dimenet}
\begin{botherref}
\oauthor{\bsnm{Gasteiger}, \binits{J.}},
\oauthor{\bsnm{Gro{\ss}}, \binits{J.}},
\oauthor{\bsnm{G{\"u}nnemann}, \binits{S.}}:
Directional message passing for molecular graphs.
arXiv preprint arXiv:2003.03123
(2020)
\end{botherref}
\endbibitem

\bibitem[\protect\citeauthoryear{Thomas et~al.}{2018}]{thomas2018tensor}
\begin{botherref}
\oauthor{\bsnm{Thomas}, \binits{N.}},
\oauthor{\bsnm{Smidt}, \binits{T.}},
\oauthor{\bsnm{Kearnes}, \binits{S.}},
\oauthor{\bsnm{Yang}, \binits{L.}},
\oauthor{\bsnm{Li}, \binits{L.}},
\oauthor{\bsnm{Kohlhoff}, \binits{K.}},
\oauthor{\bsnm{Riley}, \binits{P.}}:
Tensor field networks: Rotation-and translation-equivariant neural networks for
  3d point clouds.
arXiv preprint arXiv:1802.08219
(2018)
\end{botherref}
\endbibitem

\bibitem[\protect\citeauthoryear{Batzner et~al.}{2022}]{batzner20223}
\begin{barticle}
\bauthor{\bsnm{Batzner}, \binits{S.}},
\bauthor{\bsnm{Musaelian}, \binits{A.}},
\bauthor{\bsnm{Sun}, \binits{L.}},
\bauthor{\bsnm{Geiger}, \binits{M.}},
\bauthor{\bsnm{Mailoa}, \binits{J.P.}},
\bauthor{\bsnm{Kornbluth}, \binits{M.}},
\bauthor{\bsnm{Molinari}, \binits{N.}},
\bauthor{\bsnm{Smidt}, \binits{T.E.}},
\bauthor{\bsnm{Kozinsky}, \binits{B.}}:
\batitle{E (3)-equivariant graph neural networks for data-efficient and
  accurate interatomic potentials}.
\bjtitle{Nature communications}
\bvolume{13}(\bissue{1}),
\bfpage{2453}
(\byear{2022})
\end{barticle}
\endbibitem

\bibitem[\protect\citeauthoryear{Satorras et~al.}{2021}]{egnn}
\begin{bchapter}
\bauthor{\bsnm{Satorras}, \binits{V.G.}},
\bauthor{\bsnm{Hoogeboom}, \binits{E.}},
\bauthor{\bsnm{Welling}, \binits{M.}}:
\bctitle{E (n) equivariant graph neural networks}.
In: \bbtitle{International Conference on Machine Learning},
pp. \bfpage{9323}--\blpage{9332}
(\byear{2021}).
\bcomment{PMLR}
\end{bchapter}
\endbibitem

\bibitem[\protect\citeauthoryear{Velickovic et~al.}{2019}]{graph_infomax}
\begin{barticle}
\bauthor{\bsnm{Velickovic}, \binits{P.}},
\bauthor{\bsnm{Fedus}, \binits{W.}},
\bauthor{\bsnm{Hamilton}, \binits{W.L.}},
\bauthor{\bsnm{Li{\`o}}, \binits{P.}},
\bauthor{\bsnm{Bengio}, \binits{Y.}},
\bauthor{\bsnm{Hjelm}, \binits{R.D.}}:
\batitle{Deep graph infomax.}
\bjtitle{ICLR (Poster)}
\bvolume{2}(\bissue{3}),
\bfpage{4}
(\byear{2019})
\end{barticle}
\endbibitem

\bibitem[\protect\citeauthoryear{Hassani and
  Khasahmadi}{2020}]{hassani2020contrastive}
\begin{bchapter}
\bauthor{\bsnm{Hassani}, \binits{K.}},
\bauthor{\bsnm{Khasahmadi}, \binits{A.H.}}:
\bctitle{Contrastive multi-view representation learning on graphs}.
In: \bbtitle{International Conference on Machine Learning},
pp. \bfpage{4116}--\blpage{4126}
(\byear{2020}).
\bcomment{PMLR}
\end{bchapter}
\endbibitem

\bibitem[\protect\citeauthoryear{Qiu et~al.}{2020}]{qiu2020gcc}
\begin{bchapter}
\bauthor{\bsnm{Qiu}, \binits{J.}},
\bauthor{\bsnm{Chen}, \binits{Q.}},
\bauthor{\bsnm{Dong}, \binits{Y.}},
\bauthor{\bsnm{Zhang}, \binits{J.}},
\bauthor{\bsnm{Yang}, \binits{H.}},
\bauthor{\bsnm{Ding}, \binits{M.}},
\bauthor{\bsnm{Wang}, \binits{K.}},
\bauthor{\bsnm{Tang}, \binits{J.}}:
\bctitle{Gcc: Graph contrastive coding for graph neural network pre-training}.
In: \bbtitle{Proceedings of the 26th ACM SIGKDD International Conference on
  Knowledge Discovery \& Data Mining},
pp. \bfpage{1150}--\blpage{1160}
(\byear{2020})
\end{bchapter}
\endbibitem

\bibitem[\protect\citeauthoryear{Hu et~al.}{2020}]{hu2019strategies}
\begin{bchapter}
\bauthor{\bsnm{Hu}, \binits{W.}},
\bauthor{\bsnm{Liu}, \binits{B.}},
\bauthor{\bsnm{Gomes}, \binits{J.}},
\bauthor{\bsnm{Zitnik}, \binits{M.}},
\bauthor{\bsnm{Liang}, \binits{P.}},
\bauthor{\bsnm{Pande}, \binits{V.}},
\bauthor{\bsnm{Leskovec}, \binits{J.}}:
\bctitle{Strategies for pre-training graph neural networks}.
In: \bbtitle{International Conference on Learning Representations (ICLR)}
(\byear{2020})
\end{bchapter}
\endbibitem

\bibitem[\protect\citeauthoryear{Hou et~al.}{2022}]{hou2022graphmae}
\begin{bchapter}
\bauthor{\bsnm{Hou}, \binits{Z.}},
\bauthor{\bsnm{Liu}, \binits{X.}},
\bauthor{\bsnm{Cen}, \binits{Y.}},
\bauthor{\bsnm{Dong}, \binits{Y.}},
\bauthor{\bsnm{Yang}, \binits{H.}},
\bauthor{\bsnm{Wang}, \binits{C.}},
\bauthor{\bsnm{Tang}, \binits{J.}}:
\bctitle{Graphmae: Self-supervised masked graph autoencoders}.
In: \bbtitle{Proceedings of the 28th ACM SIGKDD Conference on Knowledge
  Discovery and Data Mining},
pp. \bfpage{594}--\blpage{604}
(\byear{2022})
\end{bchapter}
\endbibitem

\bibitem[\protect\citeauthoryear{Kingma and Welling}{2014}]{kingma2013auto}
\begin{barticle}
\bauthor{\bsnm{Kingma}, \binits{D.P.}},
\bauthor{\bsnm{Welling}, \binits{M.}}:
\batitle{Auto-encoding variational bayes}.
\bjtitle{stat}
\bvolume{1050},
\bfpage{1}
(\byear{2014})
\end{barticle}
\endbibitem

\bibitem[\protect\citeauthoryear{Zhou et~al.}{2023}]{zhou2023uni}
\begin{botherref}
\oauthor{\bsnm{Zhou}, \binits{G.}},
\oauthor{\bsnm{Gao}, \binits{Z.}},
\oauthor{\bsnm{Ding}, \binits{Q.}},
\oauthor{\bsnm{Zheng}, \binits{H.}},
\oauthor{\bsnm{Xu}, \binits{H.}},
\oauthor{\bsnm{Wei}, \binits{Z.}},
\oauthor{\bsnm{Zhang}, \binits{L.}},
\oauthor{\bsnm{Ke}, \binits{G.}}:
Uni-mol: A universal 3d molecular representation learning framework
(2023)
\end{botherref}
\endbibitem

\bibitem[\protect\citeauthoryear{Wang et~al.}{2022}]{contrast}
\begin{barticle}
\bauthor{\bsnm{Wang}, \binits{Y.}},
\bauthor{\bsnm{Wang}, \binits{J.}},
\bauthor{\bsnm{Cao}, \binits{Z.}},
\bauthor{\bsnm{Barati~Farimani}, \binits{A.}}:
\batitle{Molecular contrastive learning of representations via graph neural
  networks}.
\bjtitle{Nature Machine Intelligence}
\bvolume{4}(\bissue{3}),
\bfpage{279}--\blpage{287}
(\byear{2022})
\end{barticle}
\endbibitem

\bibitem[\protect\citeauthoryear{St{\"a}rk et~al.}{2022}]{stark20223d}
\begin{bchapter}
\bauthor{\bsnm{St{\"a}rk}, \binits{H.}},
\bauthor{\bsnm{Beaini}, \binits{D.}},
\bauthor{\bsnm{Corso}, \binits{G.}},
\bauthor{\bsnm{Tossou}, \binits{P.}},
\bauthor{\bsnm{Dallago}, \binits{C.}},
\bauthor{\bsnm{G{\"u}nnemann}, \binits{S.}},
\bauthor{\bsnm{Li{\`o}}, \binits{P.}}:
\bctitle{3d infomax improves gnns for molecular property prediction}.
In: \bbtitle{International Conference on Machine Learning},
pp. \bfpage{20479}--\blpage{20502}
(\byear{2022}).
\bcomment{PMLR}
\end{bchapter}
\endbibitem

\bibitem[\protect\citeauthoryear{Zhang et~al.}{2022}]{zhang2022dpa}
\begin{botherref}
\oauthor{\bsnm{Zhang}, \binits{D.}},
\oauthor{\bsnm{Bi}, \binits{H.}},
\oauthor{\bsnm{Dai}, \binits{F.-Z.}},
\oauthor{\bsnm{Jiang}, \binits{W.}},
\oauthor{\bsnm{Zhang}, \binits{L.}},
\oauthor{\bsnm{Wang}, \binits{H.}}:
Dpa-1: Pretraining of attention-based deep potential model for molecular
  simulation.
arXiv preprint arXiv:2208.08236
(2022)
\end{botherref}
\endbibitem

\bibitem[\protect\citeauthoryear{Wang et~al.}{2023}]{denoise}
\begin{botherref}
\oauthor{\bsnm{Wang}, \binits{Y.}},
\oauthor{\bsnm{Xu}, \binits{C.}},
\oauthor{\bsnm{Li}, \binits{Z.}},
\oauthor{\bsnm{Farimani}, \binits{A.B.}}:
Denoise pre-training on non-equilibrium molecules for accurate and transferable
  neural potentials.
arXiv preprint arXiv:2303.02216
(2023)
\end{botherref}
\endbibitem

\bibitem[\protect\citeauthoryear{Chanussot et~al.}{2021}]{chanussot2021open}
\begin{barticle}
\bauthor{\bsnm{Chanussot}, \binits{L.}},
\bauthor{\bsnm{Das}, \binits{A.}},
\bauthor{\bsnm{Goyal}, \binits{S.}},
\bauthor{\bsnm{Lavril}, \binits{T.}},
\bauthor{\bsnm{Shuaibi}, \binits{M.}},
\bauthor{\bsnm{Riviere}, \binits{M.}},
\bauthor{\bsnm{Tran}, \binits{K.}},
\bauthor{\bsnm{Heras-Domingo}, \binits{J.}},
\bauthor{\bsnm{Ho}, \binits{C.}},
\bauthor{\bsnm{Hu}, \binits{W.}}, \betal:
\batitle{Open catalyst 2020 (oc20) dataset and community challenges}.
\bjtitle{Acs Catalysis}
\bvolume{11}(\bissue{10}),
\bfpage{6059}--\blpage{6072}
(\byear{2021})
\end{barticle}
\endbibitem

\bibitem[\protect\citeauthoryear{Smith et~al.}{2017}]{smith2017ani}
\begin{barticle}
\bauthor{\bsnm{Smith}, \binits{J.S.}},
\bauthor{\bsnm{Isayev}, \binits{O.}},
\bauthor{\bsnm{Roitberg}, \binits{A.E.}}:
\batitle{Ani-1: an extensible neural network potential with dft accuracy at
  force field computational cost}.
\bjtitle{Chemical science}
\bvolume{8}(\bissue{4}),
\bfpage{3192}--\blpage{3203}
(\byear{2017})
\end{barticle}
\endbibitem

\bibitem[\protect\citeauthoryear{Liu et~al.}{2022}]{spherenet}
\begin{bchapter}
\bauthor{\bsnm{Liu}, \binits{Y.}},
\bauthor{\bsnm{Wang}, \binits{L.}},
\bauthor{\bsnm{Liu}, \binits{M.}},
\bauthor{\bsnm{Lin}, \binits{Y.}},
\bauthor{\bsnm{Zhang}, \binits{X.}},
\bauthor{\bsnm{Oztekin}, \binits{B.}},
\bauthor{\bsnm{Ji}, \binits{S.}}:
\bctitle{Spherical message passing for 3d molecular graphs}.
In: \bbtitle{International Conference on Learning Representations (ICLR)}
(\byear{2022})
\end{bchapter}
\endbibitem

\bibitem[\protect\citeauthoryear{Gasteiger et~al.}{2021}]{gasteiger2021gemnet}
\begin{barticle}
\bauthor{\bsnm{Gasteiger}, \binits{J.}},
\bauthor{\bsnm{Becker}, \binits{F.}},
\bauthor{\bsnm{G{\"u}nnemann}, \binits{S.}}:
\batitle{Gemnet: Universal directional graph neural networks for molecules}.
\bjtitle{Advances in Neural Information Processing Systems}
\bvolume{34},
\bfpage{6790}--\blpage{6802}
(\byear{2021})
\end{barticle}
\endbibitem

\bibitem[\protect\citeauthoryear{Sch{\"u}tt et~al.}{2021}]{painn}
\begin{bchapter}
\bauthor{\bsnm{Sch{\"u}tt}, \binits{K.}},
\bauthor{\bsnm{Unke}, \binits{O.}},
\bauthor{\bsnm{Gastegger}, \binits{M.}}:
\bctitle{Equivariant message passing for the prediction of tensorial properties
  and molecular spectra}.
In: \bbtitle{International Conference on Machine Learning},
pp. \bfpage{9377}--\blpage{9388}
(\byear{2021}).
\bcomment{PMLR}
\end{bchapter}
\endbibitem

\bibitem[\protect\citeauthoryear{Rapp{\'e} et~al.}{1992}]{rappe1992uff}
\begin{barticle}
\bauthor{\bsnm{Rapp{\'e}}, \binits{A.K.}},
\bauthor{\bsnm{Casewit}, \binits{C.J.}},
\bauthor{\bsnm{Colwell}, \binits{K.}},
\bauthor{\bsnm{Goddard~III}, \binits{W.A.}},
\bauthor{\bsnm{Skiff}, \binits{W.M.}}:
\batitle{Uff, a full periodic table force field for molecular mechanics and
  molecular dynamics simulations}.
\bjtitle{Journal of the American chemical society}
\bvolume{114}(\bissue{25}),
\bfpage{10024}--\blpage{10035}
(\byear{1992})
\end{barticle}
\endbibitem

\bibitem[\protect\citeauthoryear{He et~al.}{2022}]{he2022masked}
\begin{bchapter}
\bauthor{\bsnm{He}, \binits{K.}},
\bauthor{\bsnm{Chen}, \binits{X.}},
\bauthor{\bsnm{Xie}, \binits{S.}},
\bauthor{\bsnm{Li}, \binits{Y.}},
\bauthor{\bsnm{Doll{\'a}r}, \binits{P.}},
\bauthor{\bsnm{Girshick}, \binits{R.}}:
\bctitle{Masked autoencoders are scalable vision learners}.
In: \bbtitle{Proceedings of the IEEE/CVF Conference on Computer Vision and
  Pattern Recognition},
pp. \bfpage{16000}--\blpage{16009}
(\byear{2022})
\end{bchapter}
\endbibitem

\bibitem[\protect\citeauthoryear{Vincent et~al.}{2008}]{vincent2008extracting}
\begin{bchapter}
\bauthor{\bsnm{Vincent}, \binits{P.}},
\bauthor{\bsnm{Larochelle}, \binits{H.}},
\bauthor{\bsnm{Bengio}, \binits{Y.}},
\bauthor{\bsnm{Manzagol}, \binits{P.-A.}}:
\bctitle{Extracting and composing robust features with denoising autoencoders}.
In: \bbtitle{Proceedings of the 25th International Conference on Machine
  Learning},
pp. \bfpage{1096}--\blpage{1103}
(\byear{2008})
\end{bchapter}
\endbibitem

\bibitem[\protect\citeauthoryear{Chmiela et~al.}{2017}]{chmiela2017machine}
\begin{barticle}
\bauthor{\bsnm{Chmiela}, \binits{S.}},
\bauthor{\bsnm{Tkatchenko}, \binits{A.}},
\bauthor{\bsnm{Sauceda}, \binits{H.E.}},
\bauthor{\bsnm{Poltavsky}, \binits{I.}},
\bauthor{\bsnm{Sch{\"u}tt}, \binits{K.T.}},
\bauthor{\bsnm{M{\"u}ller}, \binits{K.-R.}}:
\batitle{Machine learning of accurate energy-conserving molecular force
  fields}.
\bjtitle{Science advances}
\bvolume{3}(\bissue{5}),
\bfpage{1603015}
(\byear{2017})
\end{barticle}
\endbibitem

\bibitem[\protect\citeauthoryear{Fu et~al.}{2022}]{fu2022forces}
\begin{botherref}
\oauthor{\bsnm{Fu}, \binits{X.}},
\oauthor{\bsnm{Wu}, \binits{Z.}},
\oauthor{\bsnm{Wang}, \binits{W.}},
\oauthor{\bsnm{Xie}, \binits{T.}},
\oauthor{\bsnm{Keten}, \binits{S.}},
\oauthor{\bsnm{Gomez-Bombarelli}, \binits{R.}},
\oauthor{\bsnm{Jaakkola}, \binits{T.}}:
Forces are not enough: Benchmark and critical evaluation for machine learning
  force fields with molecular simulations.
arXiv preprint arXiv:2210.07237
(2022)
\end{botherref}
\endbibitem

\bibitem[\protect\citeauthoryear{Ramakrishnan
  et~al.}{2014}]{ramakrishnan2014quantum}
\begin{barticle}
\bauthor{\bsnm{Ramakrishnan}, \binits{R.}},
\bauthor{\bsnm{Dral}, \binits{P.O.}},
\bauthor{\bsnm{Rupp}, \binits{M.}},
\bauthor{\bsnm{Von~Lilienfeld}, \binits{O.A.}}:
\batitle{Quantum chemistry structures and properties of 134 kilo molecules}.
\bjtitle{Scientific data}
\bvolume{1}(\bissue{1}),
\bfpage{1}--\blpage{7}
(\byear{2014})
\end{barticle}
\endbibitem

\bibitem[\protect\citeauthoryear{Thompson et~al.}{2022}]{thompson2022lammps}
\begin{barticle}
\bauthor{\bsnm{Thompson}, \binits{A.P.}},
\bauthor{\bsnm{Aktulga}, \binits{H.M.}},
\bauthor{\bsnm{Berger}, \binits{R.}},
\bauthor{\bsnm{Bolintineanu}, \binits{D.S.}},
\bauthor{\bsnm{Brown}, \binits{W.M.}},
\bauthor{\bsnm{Crozier}, \binits{P.S.}},
\bauthor{\bsnm{Veld}, \binits{P.J.}},
\bauthor{\bsnm{Kohlmeyer}, \binits{A.}},
\bauthor{\bsnm{Moore}, \binits{S.G.}},
\bauthor{\bsnm{Nguyen}, \binits{T.D.}}, \betal:
\batitle{Lammps-a flexible simulation tool for particle-based materials
  modeling at the atomic, meso, and continuum scales}.
\bjtitle{Computer Physics Communications}
\bvolume{271},
\bfpage{108171}
(\byear{2022})
\end{barticle}
\endbibitem

\bibitem[\protect\citeauthoryear{Jorgensen
  et~al.}{1983}]{jorgensen1983comparison}
\begin{barticle}
\bauthor{\bsnm{Jorgensen}, \binits{W.L.}},
\bauthor{\bsnm{Chandrasekhar}, \binits{J.}},
\bauthor{\bsnm{Madura}, \binits{J.D.}},
\bauthor{\bsnm{Impey}, \binits{R.W.}},
\bauthor{\bsnm{Klein}, \binits{M.L.}}:
\batitle{Comparison of simple potential functions for simulating liquid water}.
\bjtitle{The Journal of chemical physics}
\bvolume{79}(\bissue{2}),
\bfpage{926}--\blpage{935}
(\byear{1983})
\end{barticle}
\endbibitem

\bibitem[\protect\citeauthoryear{Perdew et~al.}{1996}]{perdew1996generalized}
\begin{barticle}
\bauthor{\bsnm{Perdew}, \binits{J.P.}},
\bauthor{\bsnm{Burke}, \binits{K.}},
\bauthor{\bsnm{Ernzerhof}, \binits{M.}}:
\batitle{Generalized gradient approximation made simple}.
\bjtitle{Physical review letters}
\bvolume{77}(\bissue{18}),
\bfpage{3865}
(\byear{1996})
\end{barticle}
\endbibitem

\bibitem[\protect\citeauthoryear{Bl{\"o}chl}{1994}]{blochl1994projector}
\begin{barticle}
\bauthor{\bsnm{Bl{\"o}chl}, \binits{P.E.}}:
\batitle{Projector augmented-wave method}.
\bjtitle{Physical review B}
\bvolume{50}(\bissue{24}),
\bfpage{17953}
(\byear{1994})
\end{barticle}
\endbibitem

\bibitem[\protect\citeauthoryear{Loshchilov and Hutter}{2018}]{adamw}
\begin{bchapter}
\bauthor{\bsnm{Loshchilov}, \binits{I.}},
\bauthor{\bsnm{Hutter}, \binits{F.}}:
\bctitle{Decoupled weight decay regularization}.
In: \bbtitle{International Conference on Learning Representations}
(\byear{2018})
\end{bchapter}
\endbibitem

\end{thebibliography}


\clearpage
\end{document}